\documentclass[aps,prd,twocolumn,nofootinbib,superscriptaddress]{revtex4-2}

\usepackage{amsmath,amssymb,bm,graphicx,hyperref}

\newcommand{\GeV}{\,\mathrm{GeV}}
\newcommand{\keV}{\,\mathrm{keV}}
\newcommand{\cm}{\,\mathrm{cm}}

\newcommand{\dd}{\mathrm{d}}

\begin{document}

\title{Dark Matter as the \boldmath{$Z_2$} Partner of the Standard Model Higgs Boson}

\author{Yasunori Nomura}
\affiliation{Leinweber Institute for Theoretical Physics, Department of Physics,
University of California, Berkeley, CA 94720, USA}
\affiliation{Theoretical Physics Group, Lawrence Berkeley National Laboratory,
Berkeley, CA 94720, USA}
\affiliation{RIKEN Center for Interdisciplinary Theoretical and Mathematical
Sciences (iTHEMS), RIKEN, Wako 351-0198, Japan}
\affiliation{Kavli Institute for the Physics and Mathematics of the Universe
(WPI), UTIAS, The University of Tokyo, Kashiwa, Chiba 277-8583, Japan}

%\date{September 2026}

\begin{abstract}
A possible gamma-ray excess from the Galactic halo favors dark matter with a mass of several hundred GeV. In previous work, we found that an electroweak scalar doublet is a particularly economical, and nearly uniquely selected, interpretation among simple renormalizable thermal dark-matter models. We point out that its unbroken $Z_2$ symmetry is exactly equivalent to an interchange symmetry between two Higgs doublets, making dark matter the $Z_2$-odd partner of the Standard Model Higgs boson and suggesting that the two doublets may share a common weak-scale origin.

We connect this model to the recently reported $248~\keV$ nuclear-recoil event in the LUX-ZEPLIN experiment. For dark-matter masses of $440$--$600~\GeV$, a technically natural, mass-dependent neutral-state splitting of order $350~\keV$ defines a narrow strip near the kinematic boundary where the fixed off-diagonal $Z$ interaction can produce the observed recoil. Because the scattering probes the extreme high-velocity tail of the Galactic distribution, this interpretation predicts a large annual modulation, with future events preferentially occurring around early summer. Neither observation is yet conclusive, but their common interpretation is sharply testable.
\end{abstract}

\maketitle

\section{Introduction}

The particle identity of dark matter remains unknown. A particularly compelling possibility is a weakly interacting massive particle (WIMP) whose abundance is determined by thermal freeze-out.  This framework is predictive:\ for weak-scale interactions the observed abundance points to masses near the electroweak scale, while the same interactions can be tested through indirect and direct detection.

An analysis of fifteen years of Fermi-LAT data has reported a spatially extended, approximately spherical component of Galactic halo emission with a spectral maximum near $20\GeV$~\cite{Totani:2025fxx}. Its interpretation as dark-matter annihilation favors a mass of roughly $400$--$800\,\mathrm{GeV}$, depending on the final state and halo assumptions. A subsequent independent pixel-level analysis found support for the spectral and spatial feature~\cite{Stenhouse:2026nve}. The significance of the excess and its tension with dwarf-spheroidal limits remain sensitive to foregrounds, halo profiles, substructure, and possible velocity dependence. It is therefore premature to identify the excess with dark matter. Nevertheless, the inferred mass, annihilation rate, and final states provide nontrivial information with which to test particle models.

In Ref.~\cite{Nomura:2026ntp}, we asked which simple renormalizable thermal dark-matter models could reproduce these properties while satisfying direct detection and collider bounds. A singlet scalar with a Higgs portal is excluded in the relevant regime, while an electroweak scalar doublet---the inert-doublet model~\cite{Deshpande:1977rw,Barbieri:2006dq,LopezHonorez:2006gr}---emerges as a particularly economical possibility. The gamma-ray spectrum has a best-fit mass near $440~\GeV$, with a large uncertainty, while thermal freeze-out through gauge annihilation favors approximately $500$--$600~\GeV$. The model is thus not chosen merely to fit a gamma-ray spectrum; it follows from combining the inferred signal with thermal abundance, renormalizability, experimental constraints, and minimal field content.

The LUX-ZEPLIN (LZ) Collaboration has recently reported a search extending the nuclear-recoil window to approximately $270\keV$~\cite{LZ:2026}. One event, LZ230616, is compatible with a nuclear recoil of
\begin{equation}
  E_R = 248 \pm 23_{\rm stat} \pm 23_{\rm sys} \keV,
\end{equation}
in a region with a very small known-background expectation. Across the models tested by LZ, the largest local significance is $3.4\sigma$, reduced to $2.6\sigma$ globally. A single event cannot establish new physics, especially in a newly opened extreme region of detector response. It is nonetheless notable that precisely the electroweak-doublet model selected by combining the Galactic halo excess with thermal abundance and particle-physics constraints has the structure required to produce such an event:\ the neutral scalar scatters inelastically through a $Z$ boson into a nearly degenerate neutral partner~\cite{Tucker-Smith:2001myb,Nomura:2026ntp}.

In this paper we develop this possible connection. We first explain how the usual inert-doublet model can be viewed as a model with an {\it unbroken} $Z_2$ symmetry interchanging two Higgs doublets, making dark matter a partner of the Standard Model (SM) Higgs boson. While this viewpoint is merely a change of field basis within the low-energy effective theory, it can be significant in a microscopic ultraviolet completion. If the two exchanged doublets share a common weak-scale origin---for example, if both are localized on the infrared brane of a Randall–Sundrum geometry~\cite{Randall:1999ee}---the second doublet introduces no hierarchy parametrically larger than that already associated with the SM Higgs boson. The origin of the weak scale itself may then be attributed, for example, to environmental selection~\cite{Agrawal:1997gf}.

We then show that the mass range favored by the halo signal and thermal freeze-out selects a narrow, correlated strip of neutral-state splittings near $350~\keV$, close to the kinematic boundary at which the fixed weak neutral-current interaction can yield the LZ event. Finally, we emphasize a prospective prediction of our interpretation:\ for a smooth standard halo, future high-energy recoils should exhibit a large annual modulation and cluster near early summer.

\section{Dark Matter as a {\boldmath $Z_2$} Partner of the SM Higgs Boson}

Consider the SM Higgs doublet $H$ and an additional scalar doublet $\Phi$, both with hypercharge $1/2$. In the conventional inert basis~\cite{Deshpande:1977rw,Barbieri:2006dq,LopezHonorez:2006gr}, an exact parity acts as
\begin{equation}
  H \longrightarrow H,
\qquad
  \Phi \longrightarrow -\Phi,
\label{eq:inert-parity}
\end{equation}
with all SM fields even. The most general renormalizable scalar potential consistent with this symmetry is
\begin{align}
  V ={}& m_H^2 |H|^2 + m_\Phi^2 |\Phi|^2 +\lambda_1 |H|^4 + \lambda_2 |\Phi|^4 +\lambda_3 |H|^2 |\Phi|^2
\nonumber\\
  & + \lambda_4 |H^\dagger\Phi|^2 + \frac{\lambda_5}{2} \bigl[(H^\dagger\Phi)^2 + \mathrm{h.c.}\bigr].
\label{eq:potential}
\end{align}
We take $\langle H \rangle \ne 0$ and $\langle \Phi \rangle = 0$, so the lightest neutral component of $\Phi$ is stable.

The interpretation of Eq.~(\ref{eq:inert-parity}) becomes more transparent after the basis transformation
\begin{equation}
  H_1 = \frac{H+\Phi}{\sqrt2},\qquad
  H_2 = \frac{H-\Phi}{\sqrt2}.
\label{eq:basis}
\end{equation}
The same $Z_2$ now acts as
\begin{equation}
  H_1 \leftrightarrow H_2.
\label{eq:interchange}
\end{equation}
The SM Yukawa interactions, which involve only $H$ in the inert basis, couple the fermions to the even combination $(H_1 + H_2)/\sqrt{2}$ in this basis and therefore also respect the interchange symmetry. The gauge interactions are manifestly invariant.

Since Eq.~(\ref{eq:basis}) is an invertible field redefinition, the usual inert parity is exactly equivalent to an interchange symmetry of the full low-energy theory, operator by operator and to all orders in the effective-field-theory expansion. It is therefore not a stronger assumption and does not reduce the parameter space beyond that of the usual inert-doublet model. With the vacuum satisfying $\langle H_1 \rangle = \langle H_2 \rangle$, the interchange symmetry is unbroken, and the even and odd combinations under it are, respectively,
\begin{equation}
  H = \frac{H_1+H_2}{\sqrt2},
\qquad
 \Phi = \frac{H_1-H_2}{\sqrt2},
\end{equation}
with the SM fermions coupling only to $H$. The SM Higgs doublet and the dark-matter multiplet are thus the even and odd parity eigenstates, respectively, of a common two-doublet sector. In this precise sense, dark matter is a partner of the SM Higgs boson.

This viewpoint also suggests a simple naturalness observation. Scalar masses are sensitive to the ultraviolet scale, so introducing an unrelated elementary scalar at several hundred GeV might appear to introduce an additional hierarchy. No parametrically new hierarchy is required, however, if the two exchanged doublets share a common weak-scale origin. For example, both $H_1$ and $H_2$ may be localized near the infrared brane of a Randall--Sundrum geometry~\cite{Randall:1999ee}, where their cutoff and natural mass parameters are set by the same warped-down scale. If this common scale scans in an underlying landscape, environmental selection of the weak scale~\cite{Agrawal:1997gf} can simultaneously account for the characteristic mass scale of both doublets. The odd scalar may then naturally have a mass of order the weak scale, without requiring a hierarchy parametrically distinct from the one already associated with the SM Higgs boson. This argument does not predict the numerical ratio of the two masses; it explains why a second scalar doublet at ${\cal O}(500~\GeV)$ need not introduce an independent fine-tuning problem.

\section{Galactic Halo Gamma Rays and Thermal Abundance}

The neutral dark-matter state will be denoted by $\chi$. In the regime in which scalar interactions are subdominant, its annihilation and coannihilation rates are controlled primarily by electroweak gauge interactions. At freeze-out, their characteristic size is
\begin{equation}
  \langle \sigma v\rangle_{\rm fo} \sim \frac{g^4}{{\cal O}(10)\pi m_\chi^2},
\end{equation}
where the numerical coefficient includes the contributions and statistical weights of the nearly degenerate neutral and charged states. Reproducing the observed thermal abundance then selects approximately
\begin{equation}
  m_\chi \simeq 500\text{--}600~\GeV,
\label{eq:thermal-mass-window}
\end{equation}
with the precise value depending on the charged-neutral mass splittings and scalar couplings~\cite{Banerjee:2019luv,Nomura:2026ntp}.

The Galactic halo spectrum provides an independent, although considerably less precise, determination of the mass. For the electroweak-gauge-boson final states relevant to the scalar-doublet model, the analysis of Ref.~\cite{Nomura:2026ntp} gives a best-fit value near
\begin{equation}
  m_\chi \simeq 440~\GeV.
\label{eq:halo-best-fit}
\end{equation}
The uncertainty inferred from the spectral fit is large, however, and the range favored by the gamma-ray data extends well into the thermal window of Eq.~(\ref{eq:thermal-mass-window}). Thus, the significance of the comparison is not a precise equality of two mass determinations, but that the spectral scale of the halo excess and thermal freeze-out independently point to the same several-hundred-GeV regime. In the following, we focus primarily on $m_\chi \simeq 500$--$600~\GeV$, as selected more sharply by the thermal abundance, while allowing for the broader range suggested by the gamma-ray fit.

The annihilation-rate normalization inferred under the assumption of a smooth canonical halo can exceed the thermal value. As discussed in Ref.~\cite{Nomura:2026ntp}, this does not uniquely determine the underlying particle physics. Uncertainties in the halo profile and substructure can alter the inferred normalization, while a light mediator can generate a velocity-dependent Sommerfeld enhancement. These issues are important in comparisons with dwarf-spheroidal limits, but do not affect the observation central to this paper:\ the gamma-ray spectral shape and thermal freeze-out both favor an electroweak scalar doublet in the several-hundred-GeV mass range.

\section{Inelastic scattering at LZ}

After a phase redefinition of $\Phi$, we take $\lambda_5$ to be real and write the neutral component as
\begin{equation}
  \Phi^0 = \frac{1}{\sqrt2}(\chi+i\chi'),
\end{equation}
where $\chi$ is the lighter state. Writing $v_H\equiv\langle H^0\rangle\simeq174\GeV$, the $\lambda_5$
interaction gives
\begin{equation}
  \delta\equiv m_{\chi'}-m_\chi
  \simeq\frac{|\lambda_5|v_H^2}{m_\chi}.
\label{eq:splitting}
\end{equation}
The neutral current is off diagonal, so tree-level $Z$ exchange gives
\begin{equation}
  \chi + N \longrightarrow \chi' + N,
\label{eq:inelastic-process}
\end{equation}
where $N$ denotes the target nucleus. The zero-momentum per-neutron cross section is of weak strength,
\begin{equation}
  \sigma_n^Z \simeq \frac{G_F^2 \mu_n^2}{2\pi}
  \simeq 7 \times 10^{-39}\cm^2,
\label{eq:z-cross-section}
\end{equation}
where $\mu_n = m_\chi m_n/(m_\chi+m_n)$ is the dark-matter--neutron reduced mass. This cross section would be excluded by many orders of magnitude if the process were elastic. The splitting in Eq.~(\ref{eq:splitting}) can instead place the reaction near the endpoint of the Galactic velocity distribution. A closely related off-diagonal $Z$ interaction arises for Higgsino dark matter, whose detection through inelastic upscattering has been studied in Ref.~\cite{Graham:2024syw}.

For endothermic scattering---i.e., upscattering with $\delta > 0$---the minimum incident speed required to produce a recoil $E_R$ is
\begin{equation}
  v_{\min}(E_R) = \frac{1}{\sqrt{2m_NE_R}} \left(\frac{m_NE_R}{\mu_{\chi N}}+\delta\right).
\label{eq:vmin}
\end{equation}
The minimum of this function occurs at
\begin{equation}
  E_R^* = \frac{\mu_{\chi N}}{m_N}\delta,
\qquad
  v_{\min}^* = \sqrt{\frac{2\delta}{\mu_{\chi N}}},
\label{eq:sweetspot}
\end{equation}
where $m_N$ is the target-nucleus mass and $\mu_{\chi N} = m_\chi m_N/(m_\chi+m_N)$ is the corresponding reduced mass. Consequently, a splitting of a few hundred keV suppresses low-energy recoils and moves the observable spectrum into the high-energy search region~\cite{An:2025bby}.

The LZ analysis considers inelastic realizations of the standard spin-independent contact operator $O_1$, with splittings up to $350~\keV$~\cite{LZ:2026}. Its isoscalar and isovector benchmarks reach local significances of $3.3$--$3.4\sigma$ for heavy dark matter and $\delta = 350~\keV$. The $Z$ coupling is neither of these limiting choices but a fixed neutron-dominated combination proportional to the nuclear weak charge
\begin{equation}
  Q_W^{(A,Z)} = (A-Z) - (1-4\sin^2\!\theta_W)Z,
\end{equation}
where $A$ and $Z$ are the mass and atomic numbers of the target nucleus, respectively, and $\theta_W$ is the weak mixing angle. The interaction nevertheless belongs to the same spin-independent inelastic $O_1$ class, while its proton-neutron combination and overall normalization are fixed by the electroweak theory.

At a given time, the largest kinematically allowed splitting for a specified recoil energy is
\begin{equation}
  \delta_{\max}(E_R,t) = v_{\rm cut}(t) \sqrt{2 m_N E_R} -\frac{m_N E_R}{\mu_{\chi N}},
\end{equation}
where
\begin{equation}
  v_{\rm cut}(t) = v_{\rm esc} + v_{\rm lab}(t).
\end{equation}
For comparison with the time-independent signal calculation used by LZ, we adopt its fixed reference value
\begin{equation}
  v_{\rm cut}^{\rm ref} \simeq 798~\mathrm{km/s}.
\end{equation}
Using a representative xenon-nucleus mass $m_N \simeq 122~\GeV$ then gives
\begin{equation}
  \delta_{\max}^{\rm ref}(248\keV) \simeq
  \begin{cases}
    340~\keV, & m_\chi = 440~\GeV,\\
    347~\keV, & m_\chi = 500~\GeV,\\
    352~\keV, & m_\chi = 550~\GeV,\\
    357~\keV, & m_\chi = 600~\GeV.
  \end{cases}
\label{eq:deltamax}
\end{equation}
The physical boundary varies during the year through $v_{\rm lab}(t)$; the values in Eq.~(\ref{eq:deltamax}) should therefore be understood as reference values rather than annual maxima.

The large fixed cross section in Eq.~(\ref{eq:z-cross-section}) suggests that an event rate of order unity can be obtained only when the halo integral is strongly suppressed, placing the splitting close to the kinematic boundaries in Eq.~(\ref{eq:deltamax}). The target is therefore a narrow, correlated strip in the $(m_\chi,\delta)$ plane, running from approximately $(440~\GeV,340~\keV)$ to $(600~\GeV,357~\keV)$, where the lower end of the mass range corresponds to the best fit to the Galactic halo spectrum, while the $500$--$600~\GeV$ range is favored more sharply by thermal freeze-out. This should be regarded as a kinematic target, rather than as a fitted confidence region. A quantitative determination requires the isotope-dependent weak response, detector efficiency, exposure as a function of time, and the full LZ likelihood.

The required coupling is
\begin{equation}
  |\lambda_5| \sim 6 \times 10^{-6} \left(\frac{m_\chi}{550~\GeV}\right) \left(\frac{\delta}{350~\keV}\right).
\label{eq:lambda5}
\end{equation}
This small number is technically natural:\ the limit $\lambda_5 \to 0$ restores a global $U(1)$ symmetry acting on $\Phi$. Moreover, since the freeze-out temperature is $T_{\rm fo} \simeq m_\chi/(20\text{--}25)$, one has $\delta/T_{\rm fo}\sim10^{-5}$, so the splitting has a negligible effect on gauge-dominated freeze-out, coannihilation, or the present-day gamma-ray spectrum. It can nevertheless completely control terrestrial scattering, whose available kinetic energy is only of order hundreds of keV. The heavier neutral state can decay through $\chi' \to \chi \nu \bar{\nu}$ via an off-shell $Z$ boson. Although the small splitting makes this decay relatively slow on microscopic scales, it occurs well before the present epoch, so the Galactic halo consists to excellent approximation only of the lighter state $\chi$.

\section{A seasonal prediction}

The fixed reference boundary in Eq.~(\ref{eq:deltamax}) acquires a significant time dependence once the annual variation of $v_{\rm lab}(t)$ is restored. For splittings sufficiently close to this boundary, the scattering rate consequently has a distinctive seasonal dependence. Large annual modulation is a general feature of endothermic dark matter near the Galactic velocity endpoint~\cite{LZ:2026}. The differential rate is proportional to
\begin{equation}
  \frac{\dd R}{\dd E_R} \propto F_W^2(E_R)\, \eta\bigl(v_{\min}(E_R),t\bigr),
\end{equation}
where $F_W$ is the weak nuclear form factor and
\begin{equation}
  \eta(v_{\min},t) = \int_{v>v_{\min}} \frac{f_{\rm lab}(\bm v,t)}{v}\,\dd^3v.
\end{equation}
For ordinary elastic WIMPs, annual modulation is typically a small correction to the total rate. Here, by contrast, $v_{\min}$ lies near the extreme tail of the Galactic velocity distribution. A change in the laboratory speed of ${\cal O}(10~\mathrm{km/s})$ can therefore open or close much of the available phase space, producing a modulation fraction of order unity. For a smooth Standard Halo Model, the rate is maximal near June~2 and, depending on the precise splitting, may be nearly zero during winter.

Intriguingly, LZ230616 occurred on June~16, close to this expected maximum. The event was also recorded 25~minutes after a $^{57}\mathrm{Co}$ calibration source had been removed. The source was deployed on the opposite side of the detector, and LZ found no anomalous detector population or detector condition associated with it~\cite{LZ:2026}. Both the seasonal coincidence and the proximity to the calibration were recognized only after the event was observed and should not be used to increase its present statistical significance. The seasonal coincidence does, however, motivate a prospective and falsifiable test. If the event arose from the process in Eq.~(\ref{eq:inelastic-process}), future nuclear recoils in the same high-energy region should, under the Standard Halo Model, occur preferentially in late spring and early summer. After accounting for detector live time and time-dependent efficiency, the observation of several signal-like winter events, or the absence of seasonal clustering with sufficient statistics, would disfavor this interpretation. Conversely, an exposure-corrected concentration of additional events near $200$--$270~\keV$ around early summer would support an origin in high-velocity-tail dark matter rather than an unmodulated background.

The detailed time profile is sensitive to the Galactic escape speed and to streams or other non-Maxwellian structure. Once multiple events are available, their recoil energies and dates can jointly constrain the neutral-state splitting and the high-velocity phase-space distribution. The model therefore predicts a correlated distribution in recoil energy and time, rather than merely an excess in the total event count.

\section{Discussion}

The possible connection described here is unusually constrained. The Galactic halo spectrum favors dark matter in the several-hundred-GeV mass range, while thermal abundance, renormalizability, existing constraints, and minimal field content point to an electroweak scalar doublet. Once this representation is specified, gauge invariance fixes both the dominant annihilation channels and the off-diagonal $Z$ interaction. The energy of the LZ event then identifies a narrow, correlated kinematic target in the plane of dark-matter mass and neutral-state splitting. Such a splitting is technically natural and has a negligible effect on freeze-out and the Galactic gamma-ray signal, while completely controlling terrestrial scattering. If the scattering occurs near the Galactic velocity endpoint, the model further predicts a large annual modulation rather than merely accommodating an isolated recoil event.

Both experimental ingredients require caution. The origin and significance of the Galactic halo excess remain uncertain, and the LZ result consists of a single event in an extreme region of detector response. A small expectation for known backgrounds does not exclude an unmodeled detector effect. Moreover, the direct-detection rate and its annual modulation are exceptionally sensitive to the poorly measured high-velocity tail of the local dark-matter distribution. The correlated strip identified here is therefore a kinematic target, not a fitted confidence region. A quantitative test requires the isotope-dependent weak response, detector efficiencies, time-dependent exposure, and the full LZ likelihood.

A potentially important complementary constraint arises from dark-matter capture in the Sun. Although the inelastic channel is nearly closed for terrestrial halo particles, gravitational acceleration in the solar potential can reopen scattering on heavy nuclei. Subsequent annihilation into electroweak gauge bosons may then be constrained by solar-neutrino searches~\cite{Catena:2018vzc}. The resulting bound depends sensitively on whether the captured population thermalizes, and a dedicated analysis for the scalar-doublet model is beyond the scope of this work. These caveats preclude interpreting the present observations as evidence for this specific model.

The Higgs-partner interpretation itself does not rely on either excess. The inert parity is exactly equivalent, by a change of field basis, to an interchange symmetry between two Higgs doublets. A stable electroweak multiplet can therefore be understood as the odd partner of the SM Higgs doublet. If the two-doublet sector has a common weak-scale origin, the second doublet need not introduce a hierarchy parametrically distinct from that already associated with the SM Higgs boson. Thermal freeze-out then places the odd partner in the several-hundred-GeV range, where it can be tested through indirect detection, high-recoil direct detection, and collider production.

The immediate experimental tests are clear. Additional LZ exposure, together with high-recoil analyses by other xenon experiments, can determine whether LZ230616 remains an isolated event. Under the endpoint interpretation, additional nuclear recoils should occur preferentially near the same recoil energy and, for a smooth Galactic halo, be strongly concentrated around early summer. Independently, improved foreground modeling and further analyses of the Galactic halo emission can determine whether the gamma-ray feature persists. Consistent signals in these two channels would provide strong support for the picture of dark matter as a weak-scale partner of the SM Higgs boson.

\section*{Note added}

While this paper was being completed, several related studies appeared~\cite{Su:2026rwz,Fan:2026kxx,Freese:2026sga,Wu:2026nhi,Lou:2026idn}, independently exploring interpretations of the LZ event in different dark-matter scenarios.

\section*{Note added in v2}

In a paper appearing on the same day as this work, Pospelov and Ramani showed that solar capture followed by annihilation into electroweak gauge bosons places a strong IceCube constraint on inelastic Higgsino dark matter~\cite{Pospelov:2026ewn}. The same tree-level $Z$-mediated capture mechanism operates for the scalar electroweak doublet considered here and therefore potentially places a severe constraint on our proposed interpretation of the LZ event. A definitive application requires a dedicated analysis at the lower dark-matter masses relevant here, including the post-capture evolution of the long-lived excited scalar. Our preliminary examination suggests that this post-capture effect alone is unlikely to weaken the solar-neutrino constraint by orders of magnitude. However, the splitting inferred from the LZ event is sensitive to the poorly known extreme high-velocity tail of the Galactic dark-matter distribution and may be shifted toward $500~\mathrm{keV}$ or above. Whether an allowed overlap remains therefore requires a dedicated analysis. If an allowed overlap exists at $\delta$ of order $500~\mathrm{keV}$, it would correspond to $|\lambda_5|$ of order $10^{-5}$. This value is technically natural and would have a negligible effect on freeze-out and Galactic-halo annihilation. Taking the splitting above the solar-capture threshold, expected to be of order $500~\mathrm{keV}$, evades the solar constraint without affecting either the electroweak-doublet interpretation of the Galactic halo excess or the observation that inert-doublet dark matter can be viewed as the unbroken $Z_2$ partner of the SM Higgs boson.

\begin{acknowledgments}
This work was supported by the U.S. Department of Energy, Office of Science, Office of High Energy Physics under QuantISED Award DE-SC0019380 and Contract No.\ DE-AC02-05CH11231, by MEXT KAKENHI Grant No.\ JP25K00997, and by the Japan Science and Technology Agency (JST) as part of Adopting Sustainable Partnerships for Innovative Research Ecosystem (ASPIRE), Grant No.\ JPMJAP2318.

OpenAI ChatGPT and Codex (GPT-5 series, accessed August--September 2026) were used as interactive tools for literature assistance, analytical cross-checks, and manuscript editing. The research idea, physical interpretation, and scientific conclusions were developed by the author, who independently verified all references and calculations and takes full responsibility for the content.
\end{acknowledgments}


\begin{thebibliography}{99}

\bibitem{Totani:2025fxx}
T.~Totani,
``20 GeV halo-like excess of the Galactic diffuse emission and implications for dark matter annihilation,''
JCAP \textbf{11}, 080 (2025)
%doi:10.1088/1475-7516/2025/11/080
[arXiv:2507.07209 [astro-ph.HE]].

\bibitem{Stenhouse:2026nve}
T.~R.~Stenhouse, C.~Ghag and F.~F.~Deppisch,
``The 20~GeV Galactic halo excess:\ pixel-level confirmation and consistency with sub-TeV WIMP annihilation,''
arXiv:2607.08552 [astro-ph.HE].

\bibitem{Nomura:2026ntp}
Y.~Nomura and T.~Totani,
``Electroweak doublet dark matter for a galactic halo gamma-ray excess,''
arXiv:2604.05016 [hep-ph].

\bibitem{Deshpande:1977rw}
N.~G.~Deshpande and E.~Ma,
``Pattern of symmetry breaking with two Higgs doublets,''
Phys. Rev. D \textbf{18}, 2574 (1978).
%doi:10.1103/PhysRevD.18.2574

\bibitem{Barbieri:2006dq}
R.~Barbieri, L.~J.~Hall and V.~S.~Rychkov,
``Improved naturalness with a heavy Higgs:\ an alternative road to LHC physics,''
Phys. Rev. D \textbf{74}, 015007 (2006)
%doi:10.1103/PhysRevD.74.015007
[arXiv:hep-ph/0603188].

\bibitem{LopezHonorez:2006gr}
L.~Lopez Honorez, E.~Nezri, J.~F.~Oliver and M.~H.~G.~Tytgat,
``The inert doublet model:\ an archetype for dark matter,''
JCAP \textbf{02}, 028 (2007)
%doi:10.1088/1475-7516/2007/02/028
[arXiv:hep-ph/0612275 [hep-ph]].

\bibitem{LZ:2026}
J.~Aalbers et al. [LZ Collaboration],
``Search for dark matter particle interactions in an extended nuclear recoil energy window with the LUX-ZEPLIN (LZ) experiment,''
LZ preprint (2026), available at
\url{https://lz.lbl.gov/wp-content/uploads/sites/6/2026/08/LZ_Preprint_260901_Dark_Matter_EFT_Nuclear_Recoil_Search_at_Higher_Energies.pdf}.

\bibitem{Tucker-Smith:2001myb}
D.~Tucker-Smith and N.~Weiner,
``Inelastic dark matter,''
Phys. Rev. D \textbf{64}, 043502 (2001)
%doi:10.1103/PhysRevD.64.043502
[arXiv:hep-ph/0101138 [hep-ph]].

\bibitem{Randall:1999ee}
L.~Randall and R.~Sundrum,
``A Large mass hierarchy from a small extra dimension,''
Phys. Rev. Lett. \textbf{83}, 3370 (1999)
%doi:10.1103/PhysRevLett.83.3370
[arXiv:hep-ph/9905221].

\bibitem{Agrawal:1997gf}
V.~Agrawal, S.~M.~Barr, J.~F.~Donoghue and D.~Seckel,
``Viable range of the mass scale of the standard model,''
Phys. Rev. D \textbf{57}, 5480 (1998)
%doi:10.1103/PhysRevD.57.5480
[arXiv:hep-ph/9707380].

\bibitem{Banerjee:2019luv}
S.~Banerjee, F.~Boudjema, N.~Chakrabarty, G.~Chalons and H.~Sun,
``Relic density of dark matter in the inert doublet model beyond leading order:\ the heavy mass case,''
Phys. Rev. D \textbf{100}, 095024 (2019)
%doi:10.1103/PhysRevD.100.095024
[arXiv:1906.11269 [hep-ph]].

\bibitem{Graham:2024syw}
P.~W.~Graham, H.~Ramani and S.~S.~Y.~Wong,
``Enhancing direct detection of Higgsino dark matter,''
Phys. Rev. D \textbf{111}, 055030 (2025)
%doi:10.1103/PhysRevD.111.055030
[arXiv:2409.07768 [hep-ph]].

\bibitem{An:2025bby}
H.~An, F.~Gao, J.~Liu, M.~Liu, H.~Nie and C.~Xu,
``Dark matter implications from the LZ, PandaX-4T and XENONnT data,''
arXiv:2512.05850 [hep-ph].

\bibitem{Catena:2018vzc}
R.~Catena and F.~Hellstr{\"o}m,
``New constraints on inelastic dark matter from IceCube,''
JCAP \textbf{10}, 039 (2018)
%doi:10.1088/1475-7516/2018/10/039
[arXiv:1808.08082 [astro-ph.CO]].

\bibitem{Su:2026rwz}
L.~Su, J.~M.~Yang and W.~N.~Yang,
``Inelastic dark matter signature at high recoil energy in LUX-ZEPLIN and CRESST,''
arXiv:2609.01475 [hep-ph].

\bibitem{Fan:2026kxx}
J.~Fan and M.~Reece,
``Higgsino above the sea of fog,''
arXiv:2609.01504 [hep-ph].

\bibitem{Freese:2026sga}
K.~Freese and D.~P.~Theodosopoulos,
``Higgsino dark matter interpretation of the LUX-ZEPLIN 248~keV nuclear-recoil event,''
arXiv:2609.01583 [hep-ph].

\bibitem{Wu:2026nhi}
L.~Wu, Y.~Zhang and B.~Zhu,
``TeV Higgsino dark matter from LZ nuclear recoil to Fermi-LAT gamma rays,''
arXiv:2609.01590 [hep-ph].

\bibitem{Lou:2026idn}
Y.~Lou and H.~T.~Lu,
``Fermionic dark matter absorption and the high-energy event in LUX-ZEPLIN,''
arXiv:2609.01592 [hep-ph].

\bibitem{Pospelov:2026ewn}
M.~Pospelov and H.~Ramani,
``Strong constraints on Higgsino dark matter from solar capture,''
arXiv:2609.02775 [hep-ph].

\end{thebibliography}
\end{document}